\documentclass[a4paper,11pt]{article}
\pdfoutput=1 

\usepackage{jcappub} 

\usepackage[T1]{fontenc} 

\usepackage{amsmath}
\usepackage{amssymb}
\usepackage{amsfonts}
\usepackage{bm}
\usepackage{verbatim}
\usepackage{tablefootnote}
\usepackage{enumerate}
\usepackage{afterpage}
\usepackage{xcolor}
\usepackage{natbib, hyperref}

\usepackage{graphicx}
\usepackage{tabularx}
\usepackage{booktabs}
\usepackage{pifont}

\newcommand{\orcid}[1]{\href{https://orcid.org/#1}{\,\includegraphics[width=8px]{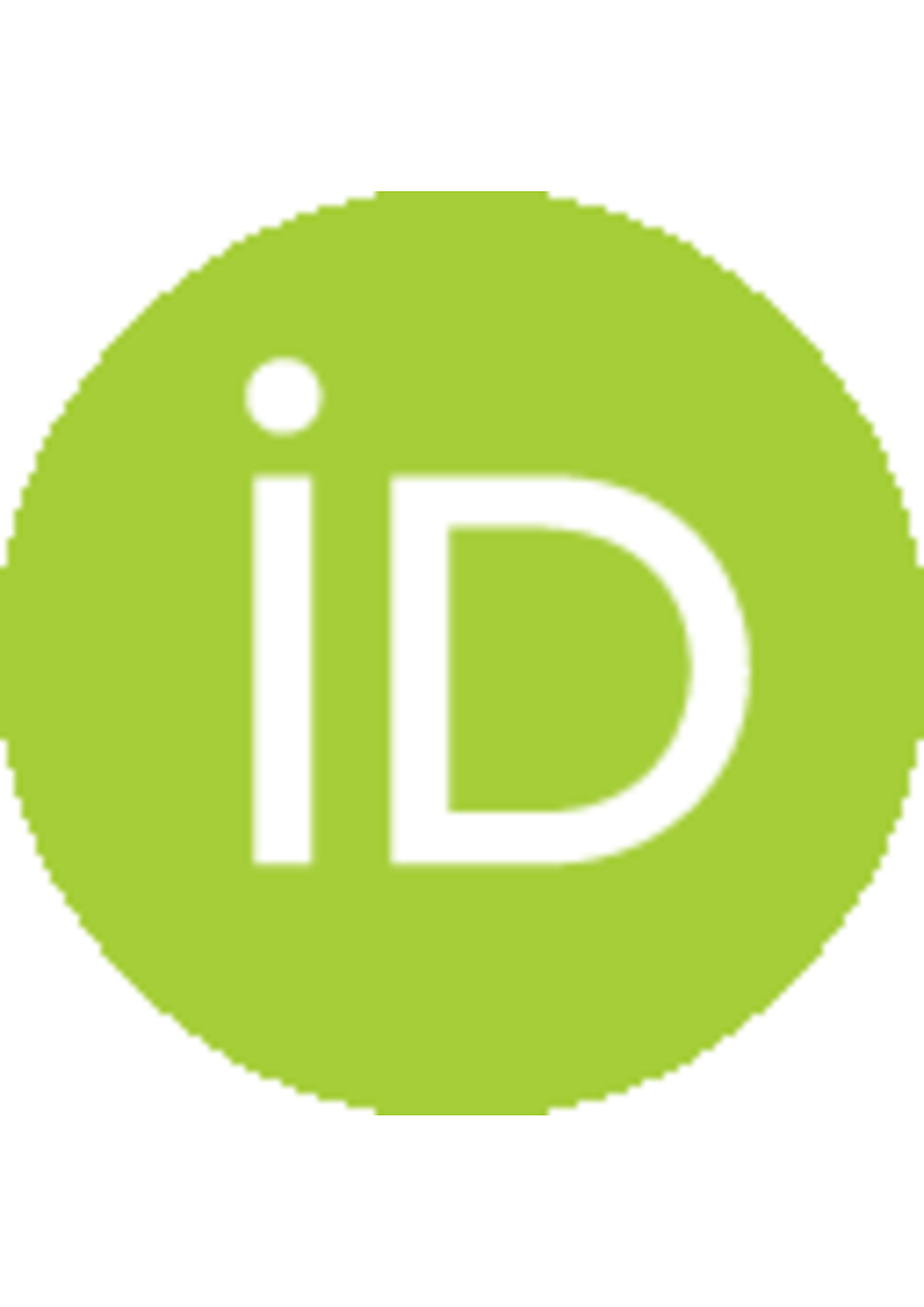}}}

\usepackage{soul}
\usepackage{xcolor}
\setstcolor{blue}

\usepackage{tikz}
\newcommand{\mathst}[1]{%
  \tikz[baseline=(X.base)]{
    \node[inner sep=0pt, outer sep=0pt] (X) {$#1$};
    \draw[blue, line width=0.5pt] (X.west) -- (X.east);
}%
}
\usepackage{etoolbox}
\robustify{\mathst}

\title{\boldmath Calibration-independent consistency test of DESI DR2 BAO and SNIa}

\author{Bikash R. Dinda$^{a,b}$\orcid{0000-0001-5432-667X},
Roy Maartens$^{a,b}$\orcid{0000-0001-9050-5894},
Chris Clarkson$^{c,a}$}

\affiliation[a]{Department of Physics \& Astronomy, University of the Western Cape, Cape Town 7535, South Africa}

\affiliation[b]{National Institute for Theoretical \& Computational Science, Cape Town 7535, South Africa}

\affiliation[c]{Department of Physics \& Astronomy, Queen Mary University of London, London E1 4NS, United Kingdom}

\emailAdd{bikashrdinda@gmail.com}
\emailAdd{rmaartens@uwc.ac.za}
\emailAdd{chris.clarkson@qmul.ac.uk}

\abstract{
We investigate the consistency between DESI DR2 BAO and three SNIa datasets, Pantheon+, Union3, and DES-Y5. Our consistency test is {calibration}-independent since it is independent of cosmological nuisance parameters such as the absolute peak magnitude $M_B$ and the comoving sound horizon at the baryon drag epoch $r_d$. Importantly, the test is also model-agnostic, independent of any model of dark energy or modified gravity. We define a tension parameter to quantify tension across different datasets compared to DESI DR2 BAO. The Pantheon+ and Union3 data have tension $\lesssim\! 1\sigma$ across their redshift ranges, whereas the DES-Y5 tension is $\gtrsim3\sigma$ near $z=1$. This hints that DES-Y5 data has significant offset values for redshifts close to 1, compared to the other SNIa datasets. Since this consistency test is independent of cosmological nuisance parameters, the tension is minimal:  other consistency tests involving differences in nuisance parameters may show greater tension.
}

\begin{document}
\maketitle
\flushbottom

\section{Introduction}

Data Release 2 (DR2) of the Dark Energy Spectroscopic Instrument (DESI) has provided unprecedented constraints on cosmic distance measurements through baryon acoustic oscillation (BAO) observations \cite{DESI:2025zgx}. Specifically, when analyzed through $w_0w_a$ parameterizations of the dark energy equation of state $w(z)$, and in combination with cosmic microwave background (CMB) \cite{Planck:2018vyg,ACT:2023kun} and type Ia supernova (SNIa) \cite{Brout:2022vxf,Rubin:2023jdq,DES:2024jxu} datasets, it gives hints of dynamical dark energy, including a possible phantom crossing ($w<-1$)  \cite{DESI:2025zgx,DESI:2025fii}.

The SNIa observations, from Pantheon+ \cite{Brout:2022vxf}, Union3 \cite{Rubin:2023jdq}, and Dark Energy Survey Year 5 (DES-Y5) \cite{DES:2024jxu}, play a crucial role in tightening constraints, especially on the low-redshift dark energy behavior and as a whole on the late-time expansion of the Universe. Dynamical dark energy would signal a major change to the standard model of cosmology. Phantom behavior of dark energy is a further potential challenge to general relativity, pointing towards the possibilities of an interacting dark sector or modified gravity (e.g. \cite{Nesseris:2025lke,Cortes:2025joz,Dinda:2025iaq,Chen:2025mlf,Guedezounme:2025wav,Leauthaud:2025azz,Wolf:2025acj,Li:2024qso,Li:2025owk,RoyChoudhury:2024wri,RoyChoudhury:2025dhe,Wang:2025znm,Ruchika:2024lgi,Hogas:2025ahb,Chaudhary:2025pcc,Chaudhary:2025vzy,Chaudhary:2025uzr,Mukherjee:2025ytj}). In light of these major possibilities, it is important in particular to check the consistency across the data sets \cite{Steinhardt:2025znn,Afroz:2025iwo}. Since tighter constraints at low redshifts are obtained from the combination of DESI DR2 BAO and SNIa (with CMB), and since both kinds of observations share a significant redshift range, consistency in these ranges must be checked \cite{Favale:2024sdq,Colgain:2024mtg,Mukherjee:2025fkf,Berti:2025phi,Scherer:2025esj,Wang:2025mqz,CosmoVerseNetwork:2025alb,Huang:2025som,Matthewson:2024ffb,Ling:2025lmw,Huang:2024gfw,Gialamas:2024lyw,Lopez-Hernandez:2025lbj}.

The consistency check between BAO and SNIa is not straightforward because they do not really measure the same observables. BAO observations provide uncalibrated cosmological distances, and to obtain the actual distance, we need to calibrate with the comoving sound horizon at the baryon drag epoch $r_d$. In order to put the supernova observables (apparent peak magnitude $m_B$) on the same footing (cosmological distance, such as the luminosity distance $d_L$), we need to calibrate the SNIa observations with the absolute peak magnitude $M_B$. This means that the results of the consistency check depend on these nuisance parameters \cite{Efstathiou:2024xcq,DES:2025tir,Yang:2025qdg,Li:2025htp,Avila:2025sjz}. Consequently, in order  to check the actual consistency, we need either to marginalize over the nuisance parameters or to consider tests that are entirely independent of the nuisance parameters.

In this analysis, we test the consistency between the DESI DR2 BAO and SNIa observations, independent of these nuisance parameters, that is, independent of the calibration. This consistency test ensures that to test consistency, we do not need any third kind of observations, such as CMB for $r_d$ or local distance observations for $M_B$.

\section{Data related observables}
\label{sec-relations}

BAO observations are related to the important Alcock–Paczynski (AP) variable,
\begin{equation}
F_{\rm AP}(z) = \frac{D_M(z)}{D_H(z)} = \frac{\widetilde{D}_M(z)}{\widetilde{D}_H(z)} \, ,
\label{eq:FAP_bao}
\end{equation}
where $D_M$ and $D_H$ are the comoving and Hubble distances, respectively, and $\widetilde{D}_M=D_M/r_d$ and $\widetilde{D}_H=D_H/r_d$ are the uncalibrated distances.
The distance modulus in SNIa observations is
\begin{equation}
\mu_B(z) = m_B(z) - M_B = 5 \log_{10} \left[ \frac{d_L(z)}{{\rm Mpc}} \right] + 25  \quad \text{where} \quad d_L(z) = (1+z)D_M(z) \, .
\label{eq:snIa}
\end{equation}
Here, $m_B$ and $M_B$ are the apparent and absolute peak magnitudes of SNIa, respectively.
We can rearrange this to express $D_M$ in terms of $m_B$ or $\mu_B$: 
\begin{equation}
D_M(z) = \alpha A(z) = \beta B(z)  \quad \text{with} \quad \alpha = {\rm e}^{- b(5+M_B)} ~ {\rm Mpc}\,, ~~ b = \frac{\ln(10)}{5}  \, , ~~ \beta = 1 ~ {\rm Gpc} \,, 
 \label{eq:mB_to_DM}
\end{equation}
and where
\begin{equation}
A(z) =(1+z)^{-1} { \exp \Big\{b\big[m_B(z)-20\big]\Big\}} \, , \quad B(z) = (1+z)^{-1}{ \exp \Big\{b\big[\mu_B(z)-40\big]\Big\}} \, .
\label{eq:main}
\end{equation}

In the flat FLRW background, $D_H=D'_M$, where a prime is a redshift derivative. Then the derivative of \autoref{eq:mB_to_DM} leads to
\begin{equation}
F_{\rm AP}(z) = \frac{A(z)}{A'(z)} = \frac{B(z)}{B'(z)} \, .
\label{eq:FAP_sn}
\end{equation}
It follows that $F_{\rm AP}$ depends only on $m_B$ or $\mu_B$ and is independent of $M_B$. 
Writing $F_{\rm AP}$ in terms of $A$ and $B$ allows us to handle the SNIa data given in $m_B$ or $\mu_B$.
It may seem unnecessary to express $F_{\rm AP}$ in new variables $A,B$ rather than $m_B,\mu _B$, where $F_{\rm AP}(z)={(1+z)}/{[b(1+z)m'_B(z)-1]}={(1+z)}{[b(1+z)\mu' _B(z)-1]}$. The reason is that there are subtle differences when we apply Gaussian Processes (GP). In 
\autoref{sec-gp_mB_vs_A}, 
we show that it is better to apply GP to $A, B$ than to $m_B, \mu _B$.

\begin{figure}[!htbp]
\centering
\includegraphics[width=0.45\linewidth]{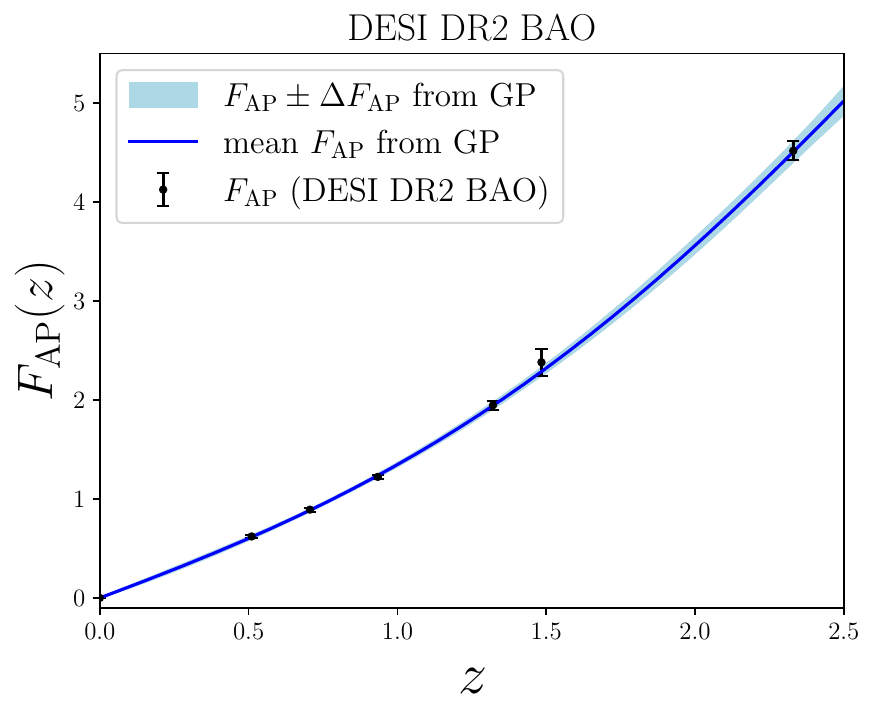}
\includegraphics[width=0.45\linewidth]{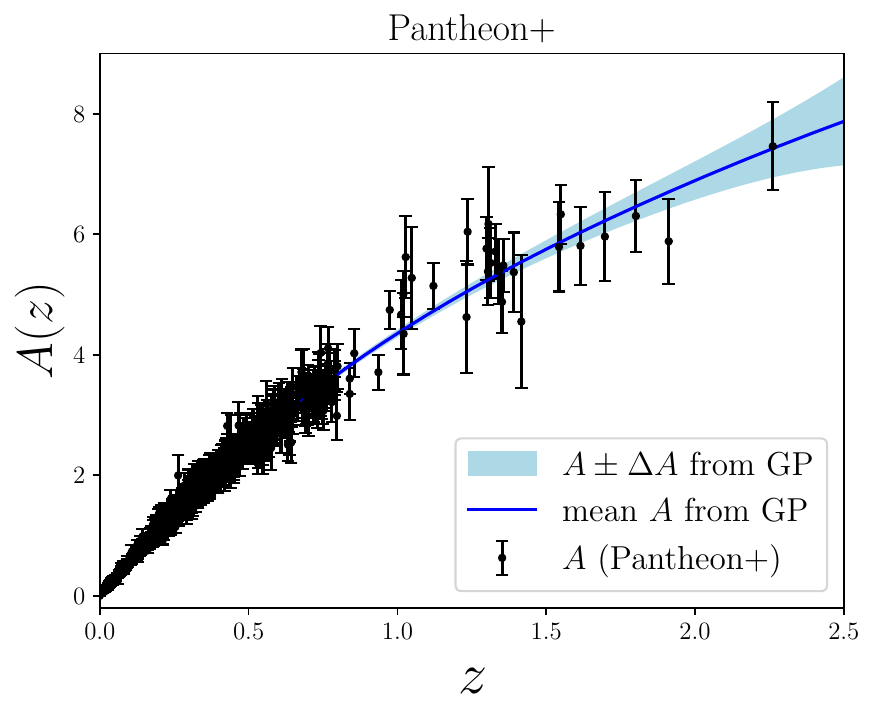} \\
\includegraphics[width=0.45\linewidth]{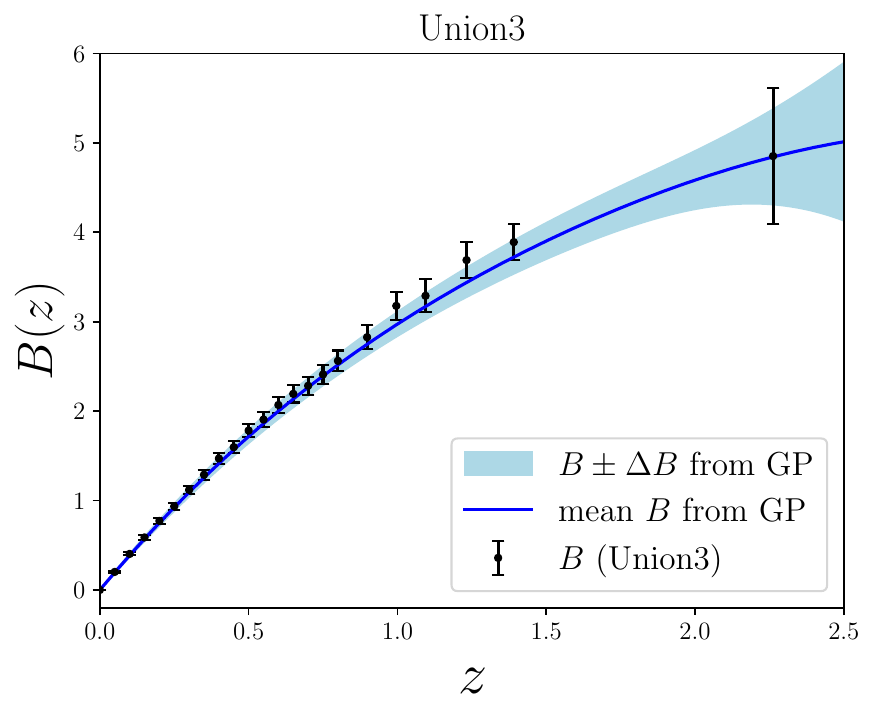}
\includegraphics[width=0.45\linewidth]{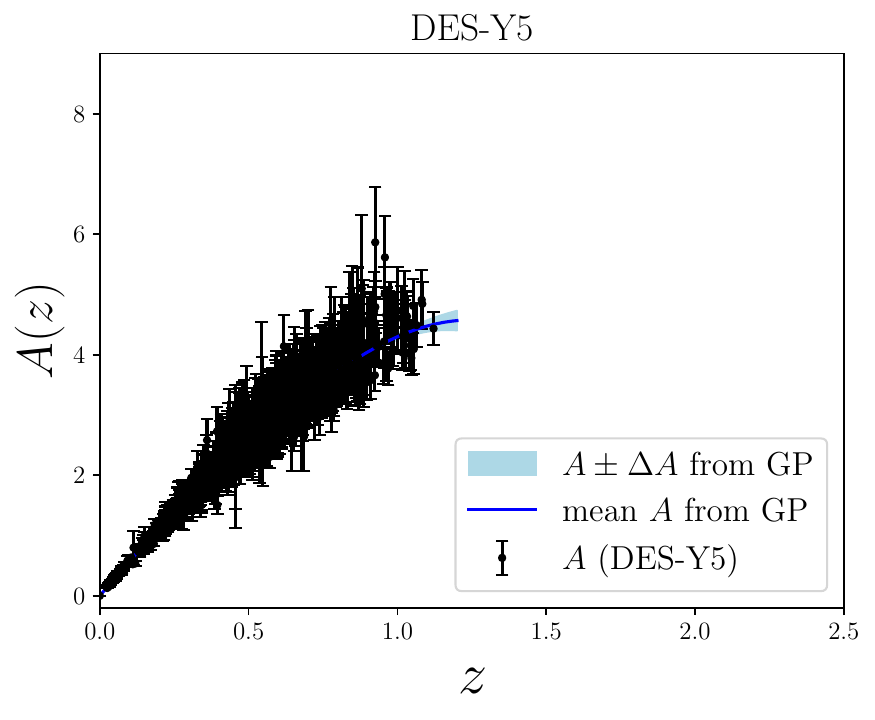}
\caption{Gaussian Process predictions (blue) compared to observational data (black), where $A$ and $B$ are defined in \autoref{eq:main}.
}
\label{fig:data_vs_gp}
\end{figure}

BAO observations from DESI DR2 directly provide data for $F_{\rm AP}$. In addition, we consider three SNIa datasets: Pantheon+, Union3, and DES-Y5. The first two have data up to redshift $z\sim2.4$, whereas the third reaches $z\sim1.2$. SNIa observations provide data of $m_B$ in Pantheon+ and DES-Y5, but Union3 has not officially released the data of apparent magnitude $m_B$, providing only binned $\mu_B$ data. As mentioned above, this is not an issue in our case, since our methodology applies to both $m_B$ and $\mu_B$.
We directly convert these SNIa data to $A$ and $B$ data, using \autoref{eq:main} with uncertainty predictions (including both self- and cross-variances). These are shown in \autoref{fig:data_vs_gp} {(black data points)}.

Note that at $z=0$, we place theoretical priors of $F_{\rm AP}(0)=0$, $A(0)=0$, $B(0)=0$ in order to have consistent theoretical values according to a FLRW metric. However, this is not strictly necessary, nor does it alter results, except when $z$ is very close to zero. We also emphasize that, if we want to check the violation of the FLRW model at $z\approx0$ (which is not the aim of this paper), then we should not consider these theoretical priors.

\section{Results}
\label{sec-result}

We apply GP to the DESI DR2 $F_{\rm AP}$ data directly in order to reconstruct a smooth function and the associated uncertainties. For this purpose, we consider a zero-mean function and a squared-exponential kernel covariance function. Similarly, we apply GP to $A$ (Pantheon+ and DES-Y5) and $B$ (Union3) data, to reconstruct these and their first derivatives. The results are shown in \autoref{fig:data_vs_gp}  for a visual test of the accuracy of GP regression predictions (blue curves) compared to the actual data (black data points).

In order to check the robustness of the results, we show in \autoref{sec-robustness} that the GP reconstructed results are compatible with 4th and 5th-order polynomial reconstructed results.

\begin{figure}[!htbp]
\centering
\includegraphics[width=0.51\linewidth , height=0.3\textheight]{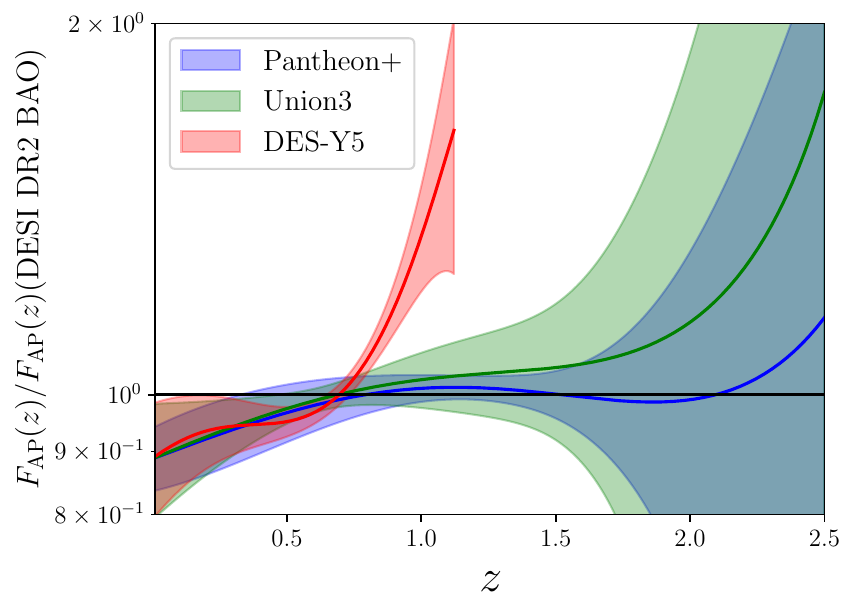}
\includegraphics[width=0.47\linewidth , height=0.3\textheight]{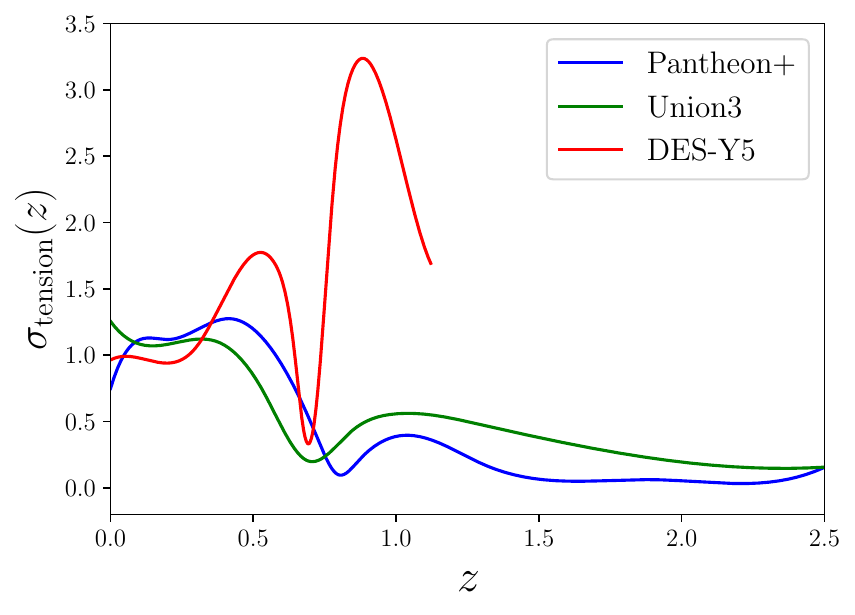}
\caption{{\em Left:} $F_{\rm AP}$ comparison of SNIa data to DESI data. {\em Right:} Tension between SNIa  and DESI datasets, as given by \autoref{eq:tension}.
}
\label{fig:FAP}
\end{figure}

Using these GP predicted smooth functions and their derivatives (wherever needed), we compute $F_{\rm AP}(z)$. We compare these to the DESI data in the left panel of \autoref{fig:FAP}, which displays the ratios of the SNIa $F_{\rm AP}$ and the DESI $F_{\rm AP}$. The shading shows the 1$\sigma$ errors on these ratios. At lower redshifts ($z<0.3-0.4$) all three SNIa datasets are  $\sim\!1\sigma$ away from the DESI DR2 BAO data. In the middle redshift range ($0.3-0.4<z<1.15$), we see that the discrepancy between DESI DR2 BAO, Pantheon+, and Union3 is decreasing to less than 1$\sigma$. However, the DES-Y5 data show a strong growth in inconsistency with DESI {DR2 BAO}.  In the highest redshift range, where there is no DES-Y5 data, we see that the inconsistency for Pantheon+ and Union3  decreases even further, well within 1$\sigma$. 

In order to properly quantify the inconsistency across different observations, we define a tension parameter in units of 1$\sigma$ confidence intervals:
\begin{equation}
\sigma_{\rm tension}(z) = \frac{\big|F_{\rm AP}(z)-F_{\rm AP}(z)(\text{\small DESI DR2 BAO})\big|}{\Big\{\big[\Delta F_{\rm AP}(z)\big]^2+\big[\Delta F_{\rm AP}(z)(\text{\small DESI DR2 BAO})\big]^2\Big\}^{1/2}} \, .
\label{eq:tension}
\end{equation}
We display this tension parameter in the right panel of \autoref{fig:FAP}. It is clear that the inconsistencies of Pantheon+ and Union3 with DESI DR2 BAO are around 1$\sigma$ for $z\lesssim 0.5$ and decrease with increasing redshift for $z\gtrsim0.5$. DES-Y5 data has similar low inconsistency for $z\lesssim0.5$, but the inconsistency increases for
$z\gtrsim0.5$, up to more than 3$\sigma$ near $z=1$. DES-Y5 data have significantly offset values near $z=1$, compared to Pantheon+ and Union3 data.

We assumed a spatially flat FLRW spacetime for these results. The relation in \autoref{eq:FAP_sn} is modified by cosmic curvature, but curvature hardly makes any change to the results in the consistency test, as shown in \autoref{sec-curvature}.

\section{Conclusions}
\label{sec-conclusion}

We presented a method for testing different datasets, which is independent of cosmological nuisance parameters, such as the peak absolute magnitude of SNIa. This consistency test is also model-agnostic.

We find DESI DR2 BAO, Pantheon+, and Union3 data are at around 1$\sigma$ tension for $z\lesssim0.5$, but decreasing for $z >0.5$.
The DES-Y5 data is consistent with the other two SNIa datasets at $z\lesssim0.5$, but notably for $0.5\lesssim z \lesssim 1.15$, the inconsistency increases, reaching more than 3$\sigma$ near $z=1$. DES-Y5 data is significantly offset in this redshift range -- and this inconsistency is independent of $M_B$. The differences in $M_B$ may even increase the inconsistency. This offset might have significant effects on the constraints on dark energy. 
We showed that the spatial curvature of the universe does not affect these conclusions. 

If the mismatch between $F_{\rm AP}$ from BAO and from SNIa (using \autoref{eq:FAP_sn})  is not due to any systematic errors, this could hint at fundamental violations, such as:  breakdown of the cosmic distance duality relation, i.e. $d_L=(1+z)^2 d_A$, where $d_A$ is the angular diameter distance;  violation of the standard candle assumptions for SNIa; violation of the FLRW geometry of spacetime or of standard cosmic physics. We briefly discuss this in \autoref{sec-consequences}.

\acknowledgments
BRD and RM are supported by the South African Radio Astronomy Observatory and the National Research Foundation (Grant No. 75415).

\appendix

\section{Gaussian Processes and $m_B, \mu_B$}
\label{sec-gp_mB_vs_A}

\begin{figure}[!htbp]
\centering
\includegraphics[width=0.45\linewidth]{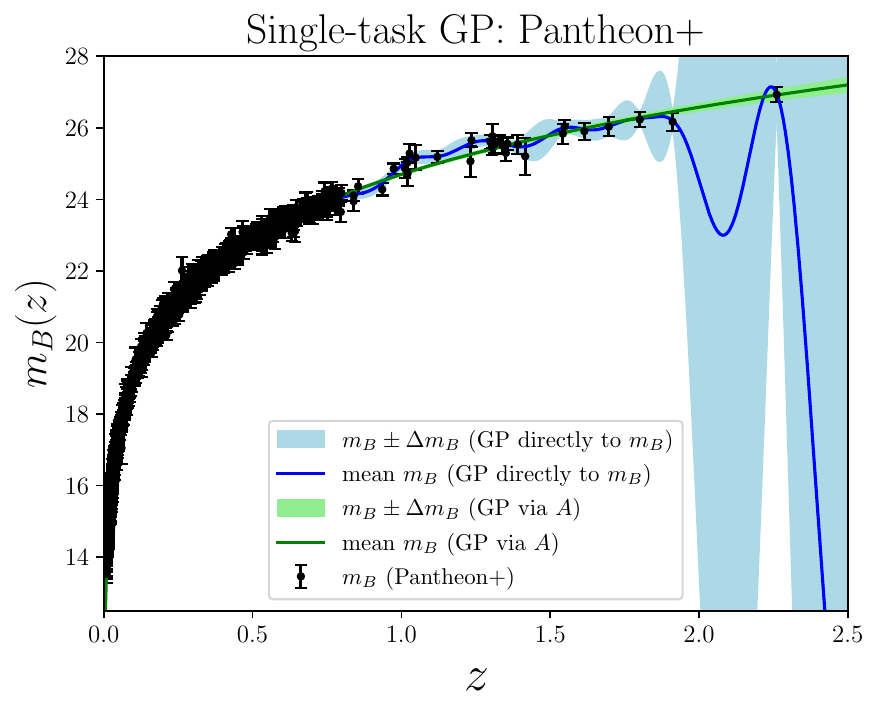}
\includegraphics[width=0.45\linewidth]{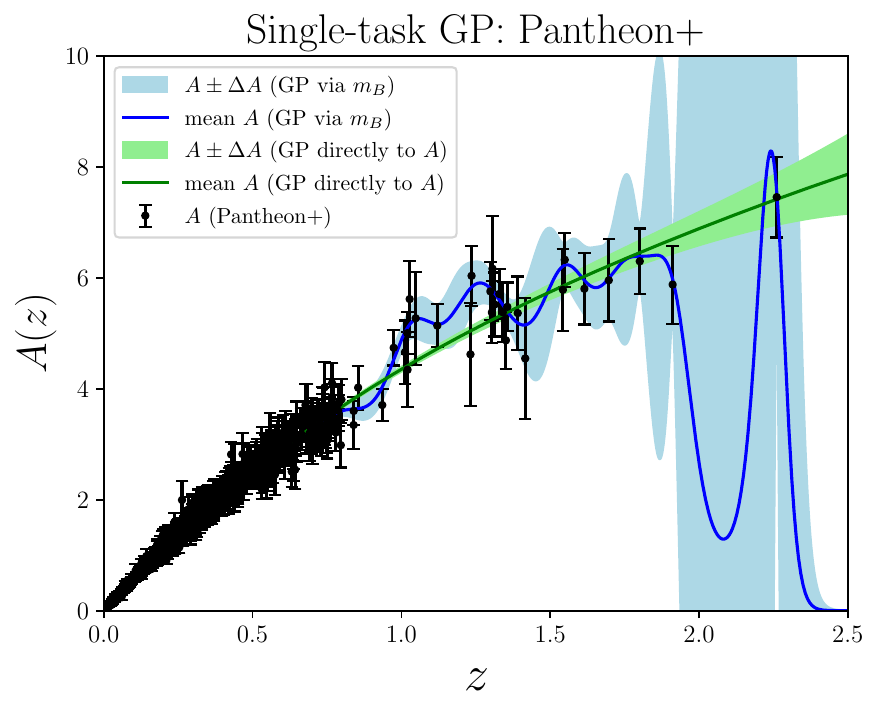} \\
\includegraphics[width=0.45\linewidth]{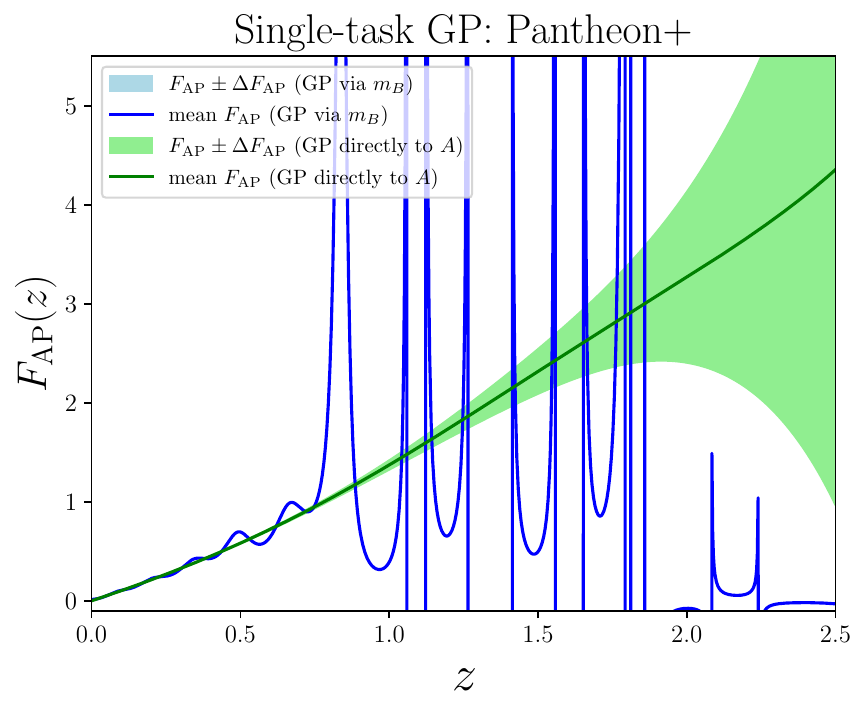}
\includegraphics[width=0.45\linewidth]{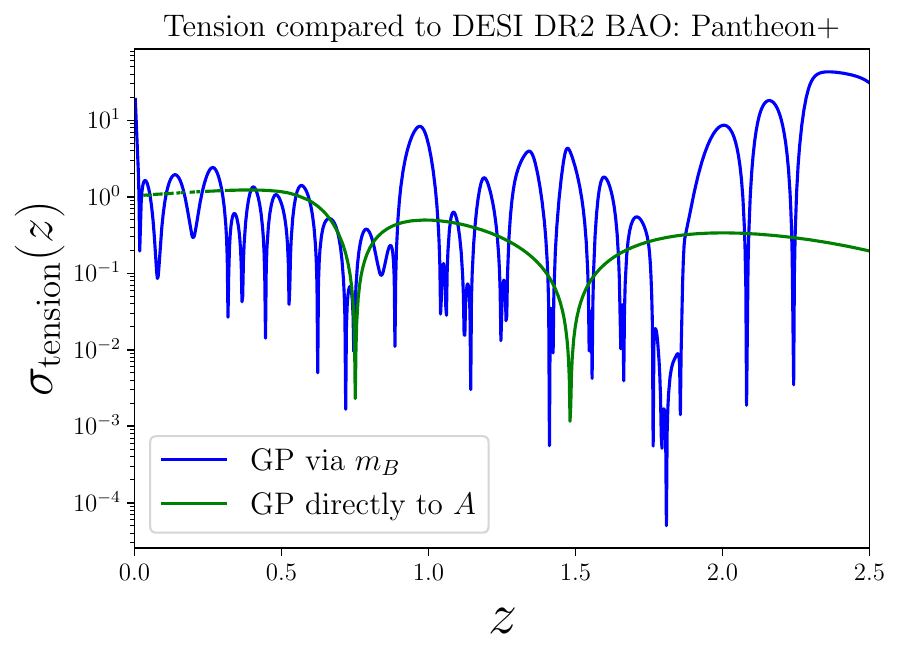}
\caption{
{GP applied to $m_B$ compared to $A$, using Pantheon+ data.}
}
\label{fig:gp_m_vs_A}
\end{figure}

We used $A, B$ instead of $m_B, \mu_B$ because the former variables lead to much smoother reconstructions. For example, the behavior of $m_B=20+\log[(1+z)A]/b$ is more nonlinear in $z$ than $A$, even though it is derived from the same data (Pantheon+ or other SNIa data). When we apply GP to $m_B$, the value of the hyperparameter of the squared-exponential kernel is $l=0.13$,  significantly lower than $l=2.44$ in the case of $A$. The lower $l$ value means the reconstructed function changes more rapidly, with oscillations. Not only $m_B$, but also its derivatives are more oscillating due to the lower $l$. This is shown in \autoref{fig:gp_m_vs_A}. Clearly better results follow from applying GP  to $A$ than to $m_B$. This is also true for DES-Y5 data and  Union3 data. Similarly, GP reconstruction of $B$ is better behaved than that of $\mu_B$. All these points apply to a zero-mean function or a mean function that is not very close to the actual true underlying model of the data. More detailed investigation of these features is beyond the scope of this paper.

The same conclusions apply for polynomial regression (see \autoref{sec-robustness}) as for GP.

\section{Robustness of the results}
\label{sec-robustness}

\begin{figure}[!htbp]
\centering
\includegraphics[width=0.45\linewidth]{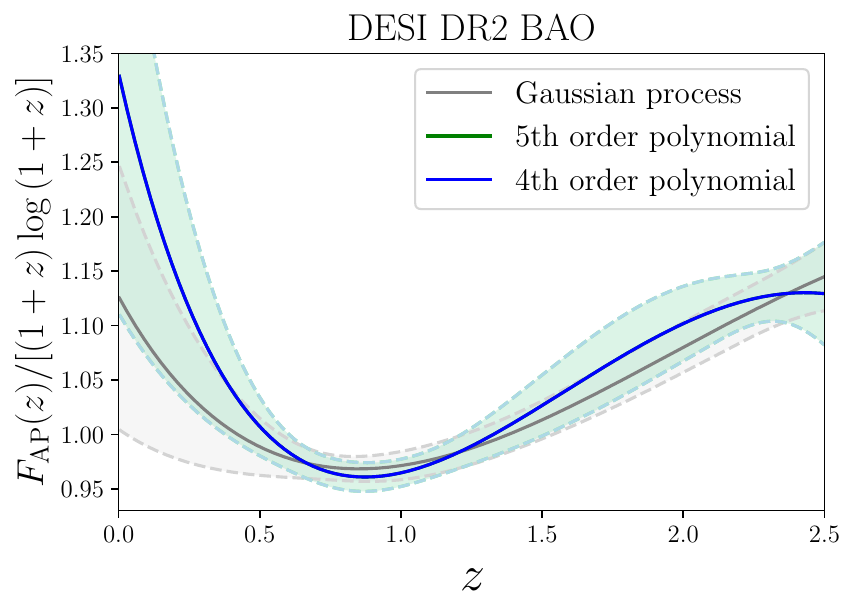}
\includegraphics[width=0.45\linewidth]{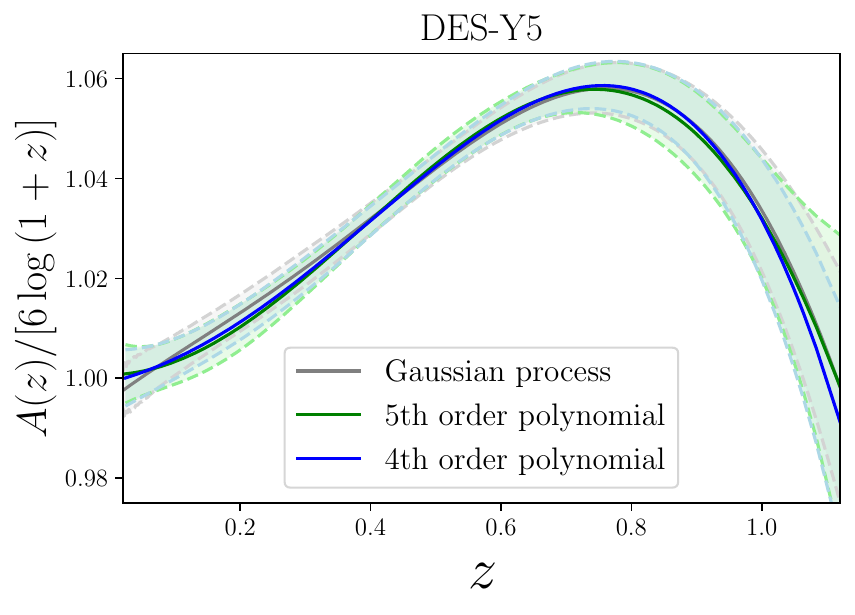}
\caption{Robustness of the results.
}
\label{fig:robustness}
\end{figure}

To test the robustness of the results, we compare the GP reconstructed values to the 4th and 5th-order polynomial regression fitted values of the main functions. We see in \autoref{fig:robustness} that the results agree with each other well within 1$\sigma$. We show this for DESI DR2 BAO (left) and DES-Y5 (right) data, but it is also true for the other two SNIa datasets.

\section{Effects of cosmic curvature in the consistency test}
\label{sec-curvature}

In curved spacetime, $\widetilde{D}'_M \neq \widetilde{D}_H$ \citep{Dinda:2022vmb}:
\begin{equation}
\widetilde{D}'_M = \widetilde{D}_H \Big[1+ \gamma \widetilde{D}_M^2(z)\Big]^{1/2}\quad \text{where} \quad  \gamma = \Omega_{\rm K0} \Big( \frac{H_0 r_d}{c} \Big)^2 \, .
\label{eq:curvature_correction}
\end{equation}
Using this and \autoref{eq:mB_to_DM}, we find the connection between SNIa and BAO observables in general in curved spacetime:
\begin{equation}
\frac{A(z)}{A'(z)} = \frac{B(z)}{B'(z)} = {F_{\rm AP}(z)}{\Big[1 + \gamma} \widetilde{D}_M^2(z)\Big]^{-1/2} \, .
\label{eq:curvature_correction_2}
\end{equation}
DESI or any BAO observations provide both $\widetilde{D}_M$ and $\widetilde{D}_H$. Then any reconstruction technique, such as GP or polynomial regression used here, which predicts derivatives ($\widetilde{D}'_M$) along with the main function ($\widetilde{D}_M$), allows us to find model-independent reconstruction of $\gamma$ using \autoref{eq:curvature_correction},  alongside the main reconstructions. We use multi-task GP, extending the methodology for nonzero cosmic curvature in \cite{Dinda:2024ktd}, to DESI DR2 BAO data, and we find that
\begin{equation}
\text{DESI DR2 BAO:} \quad \gamma = (0.5\pm 6.5) \times 10^{-5} \, . \nonumber
\label{eq:beta}
\end{equation}

\begin{figure}[!htbp]
\centering
\includegraphics[width=0.55\linewidth]{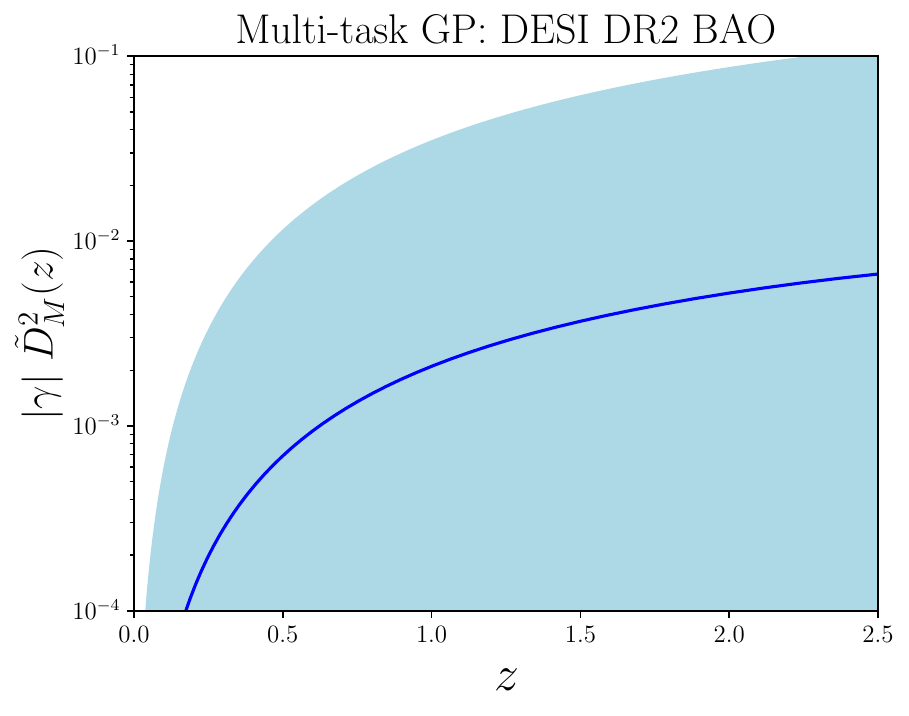}
\caption{The curvature correction term $|\gamma| \widetilde{D}_M^2(z)$.
}
\label{fig:curved}
\end{figure}

The curvature correction term $|\gamma| \widetilde{D}_M^2(z)$ is shown in \autoref{fig:curved} with associated 1$\sigma$ uncertainties. It increases from zero (at $z=0$) to a maximum $\sim 10^{-2}$ at $z=2.5$, i.e. $|\gamma| \widetilde{D}_M^2(z) \ll 1$ for this entire redshift range. Consequently, there are negligible effects of cosmic curvature in our consistency test. However, $|\gamma| \widetilde{D}_M^2(z)$ continues to increase  for higher redshifts $z>2.5$, because $\widetilde{D}_M$ increases with increasing redshift. Eventually, at a high enough redshift, the term becomes comparable to unity, and then cosmic curvature shows its effect. For this reason, curvature has effects on the CMB angular scales, but not in the setup considered here. Significant constraints on cosmic curvature come only when we include CMB data in any cosmological data analysis, e.g., as in Table~V  of~\cite{DESI:2025zgx}, unless any future low redshift observations provide a significant value of $\gamma$ comparable to $10^{-2}$ or larger.

\section{Potential consequences of the consistency test}
\label{sec-consequences}

If the mismatch between BAO and SNIa is not due to systematics, this could be a violation of standard physics. We consider two possibilities.

\subsection{Violation of cosmic distance duality relation}

\begin{figure}[!htbp]
\centering
\includegraphics[width=0.55\linewidth]{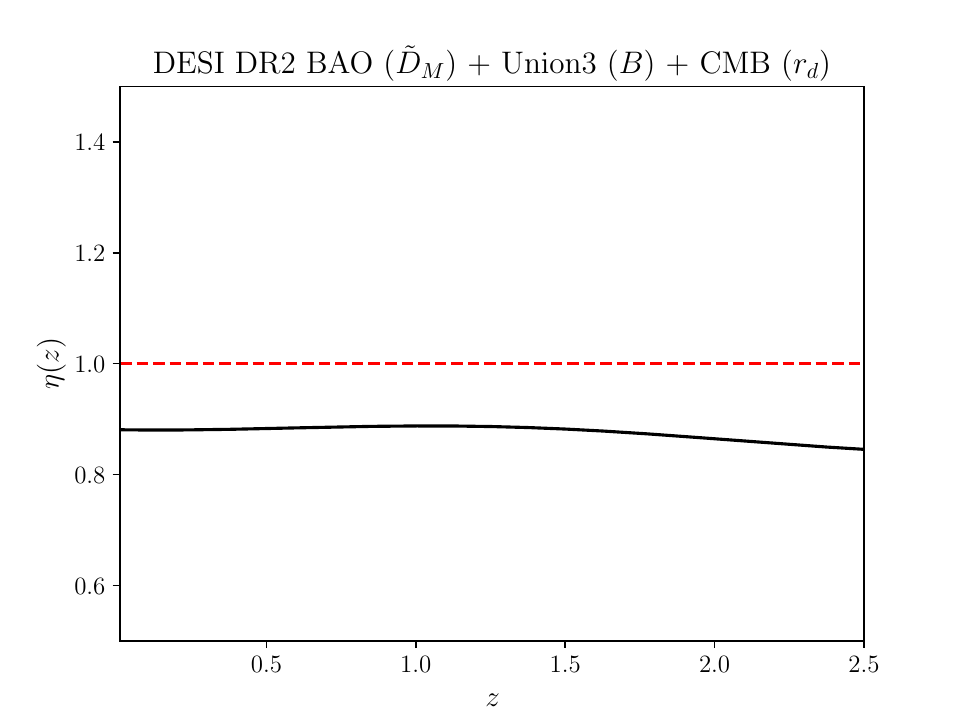}
\caption{
{$\eta$ obtained by combining DESI DR2 BAO, Union3 and CMB.}
}
\label{fig:eta}
\end{figure}

We introduce a  violation parameter $\eta$ in \autoref{eq:snIa}: 
\begin{equation}
m_B(z) = M_B + 5 \log_{10} \left[ \frac{ \eta(z) (1+z)D_M(z) }{{\rm Mpc}} \right] + 25 \, ,
\label{eq:modified_snIa}
\end{equation}
which leads to
\begin{equation}
\eta(z) = \frac{\exp \Big\{b\big[m_B(z)-M_B-25\big]\Big\}}{(1+z)D_M(z)} = \frac{\alpha A(z)}{r_d \widetilde{D}_M(z)} = \frac{\beta B(z)}{r_d \widetilde{D}_M(z)} \, ,
 \label{eq:eta}
\end{equation}
using \autoref{eq:mB_to_DM},  \autoref{eq:main} and  $\widetilde{D}_M=D_M/r_d$. This can be rewritten as
\begin{align}
\frac{r_d \eta}{\alpha} &= \frac{A}{\widetilde{D}_M} \quad \text{comparison between DESI DR2 BAO and Pantheon+ or DES-Y5,}
\label{eq:case1} \\
\frac{r_d \eta}{\beta} &= \frac{B}{\widetilde{D}_M} \quad \text{comparison between DESI DR2 BAO and Union3.}
\label{eq:case2}
\end{align}
These two equations show that $\eta$ is completely degenerate either with  $r_d$ (for Union3) or with $r_d$ and $M_B$ (for Pantheon+ or DES-Y5). In order to test the cosmic distance duality relation, we use $r_d$  from CMB observations, assuming no violation of standard physics in the early Universe. We apply this to \autoref{eq:case2}. We do not do any test for the case \autoref{eq:case1}, because for this case we need to consider $M_B$ from local observations, which will automatically include the $M_B$ tension (and consequently the Hubble tension) \cite{Dinda:2021ffa,Efstathiou:2021ocp}. In other words,  for \autoref{eq:case1}, we cannot really break the degeneracy between these tensions and the violation of cosmic distance duality. In \autoref{fig:eta}, we test this for the combination of DESI DR2 BAO, Union3 SNIa, and CMB data, and we find there is no evidence of distance duality violation. Note that here we used $r_d = (147.43 \pm 0.25)$\, Mpc,  from  Planck+ACT DR6 CMB observations \cite{Planck:2018vyg,ACT:2023kun}, which is consistent with the DESI DR2 main paper \cite{DESI:2025zgx}.

\begin{figure}[!htbp]
\centering
\includegraphics[width=0.45\linewidth]{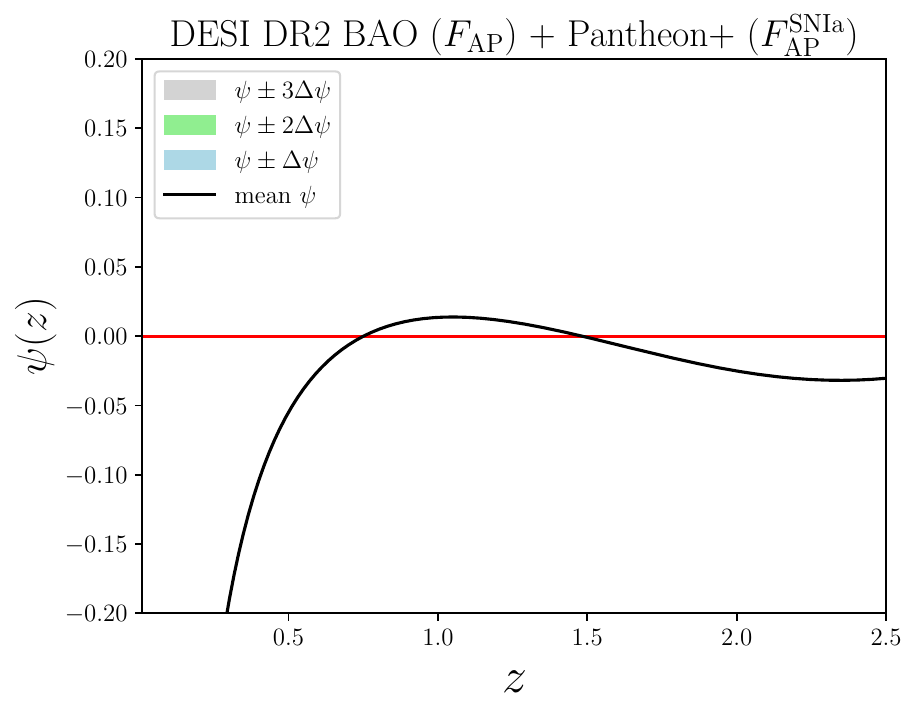}
\includegraphics[width=0.45\linewidth]{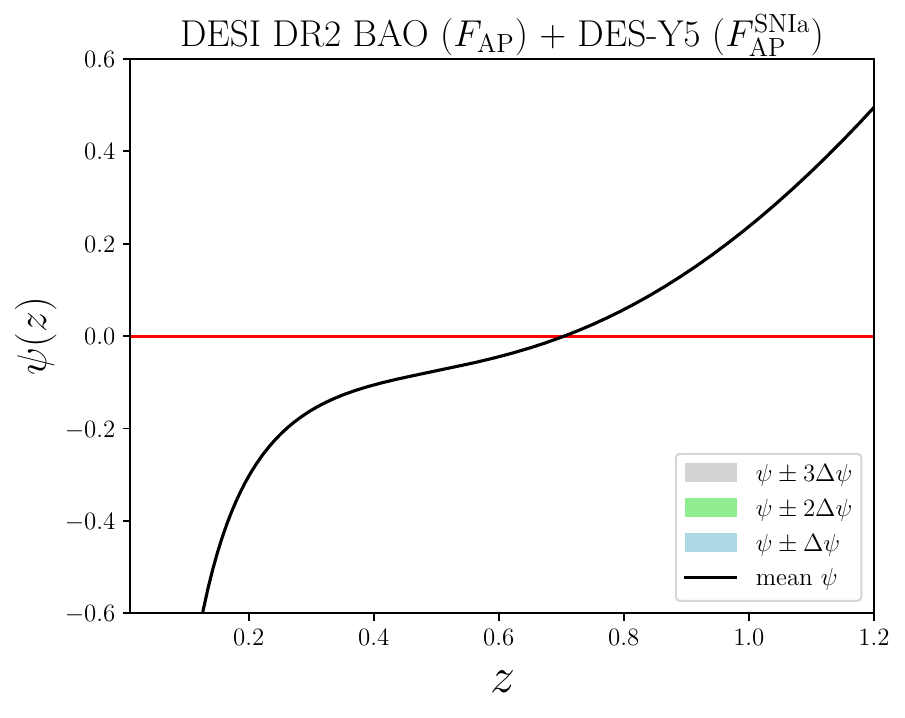}
\caption{
Test of standard candle assumptions, with no other violation of standard physics.
}
\label{fig:MB_prime}
\end{figure}

\subsection{Violation of standard candle assumptions}

Here we test whether there is any violation of the standard candle assumptions, i.e., any redshift dependence in $M_B$, which is equivalent to redshift dependence of $\alpha$ (see \autoref{eq:mB_to_DM}). We assume that distance duality holds. Rewriting \autoref{eq:mB_to_DM} as
\begin{equation}
    \alpha (z) = r_d\,\frac{ \widetilde{D}_M(z)}{A(z)} \, ,
\label{eq:alpha}
\end{equation}
we find that
\begin{equation}
    \alpha' (z) = r_d\,\frac{ \big[A(z) \widetilde{D}'_M(z) - A'(z) \widetilde{D}_M(z) \big] }{A^2(z)} \, .
\label{eq:alpha_prime}
\end{equation}
These two equations imply
\begin{equation}
    \psi \equiv \frac{\alpha'}{\alpha} = \frac{\widetilde{D}'_M}{\widetilde{D}_M} - \frac{A'}{A} = \frac{1}{F_{\rm AP}} - \frac{1}{F_{\rm AP}^{\rm SNIa}}  \qquad \text{with} \qquad F_{\rm AP}^{\rm SNIa} = \frac{A}{A'}\, .
\label{eq:psi}
\end{equation}
Violation of the standard candle assumption corresponds to $\psi \neq 0$. This test is independent of $r_d$. The test is not directly applicable to Union3 data because there is no $M_B$ parameter involved to be tested.
\autoref{fig:MB_prime} shows $\psi$ for Pantheon+ (left) and DES-Y5 (right), combined with DESI DR2 data. It is evident that for Pantheon+, with $z<0.5$, there is redshift evolution of $M_B$ at slightly more than 1$\sigma$, but well within 1$\sigma$ for $z>0.5$. For DES-Y5, it is similar for lower reshift, but at around $z=1$, there is up to 3$\sigma$ evidence for redshift evolution of $M_B$. In other words, if there is no inconsistency coming from any other modification, then this is evidence for the violation of standard candle assumptions.

\bibliographystyle{JHEP}
\bibliography{references}

\providecommand{\href}[2]{#2}\begingroup\raggedright\begin{thebibliography}{10}

\bibitem{DESI:2025zgx}
{\bf DESI} Collaboration, M.~Abdul~Karim et~al., {\it {DESI DR2 Results II:
  Measurements of Baryon Acoustic Oscillations and Cosmological Constraints}},
  \href{http://arxiv.org/abs/2503.14738}{{\tt arXiv:2503.14738}}.

\bibitem{Planck:2018vyg}
{\bf Planck} Collaboration, N.~Aghanim et~al., {\it {Planck 2018 results. VI.
  Cosmological parameters}},  {\em Astron. Astrophys.} {\bf 641} (2020) A6,
  [\href{http://arxiv.org/abs/1807.06209}{{\tt arXiv:1807.06209}}]. [Erratum:
  Astron.Astrophys. 652, C4 (2021)].

\bibitem{ACT:2023kun}
{\bf ACT} Collaboration, M.~S. Madhavacheril et~al., {\it {The Atacama
  Cosmology Telescope: DR6 Gravitational Lensing Map and Cosmological
  Parameters}},  {\em Astrophys. J.} {\bf 962} (2024), no.~2 113,
  [\href{http://arxiv.org/abs/2304.05203}{{\tt arXiv:2304.05203}}].

\bibitem{Brout:2022vxf}
D.~Brout et~al., {\it {The Pantheon+ Analysis: Cosmological Constraints}},
  {\em Astrophys. J.} {\bf 938} (2022), no.~2 110,
  [\href{http://arxiv.org/abs/2202.04077}{{\tt arXiv:2202.04077}}].

\bibitem{Rubin:2023jdq}
D.~Rubin et~al., {\it {Union Through UNITY: Cosmology with 2,000 SNe Using a
  Unified Bayesian Framework}},  {\em Astrophys. J.} {\bf 986} (2025) 231,
  [\href{http://arxiv.org/abs/2311.12098}{{\tt arXiv:2311.12098}}].

\bibitem{DES:2024jxu}
{\bf DES} Collaboration, T.~M.~C. Abbott et~al., {\it {The Dark Energy Survey:
  Cosmology Results with {\ensuremath{\sim}}1500 New High-redshift Type Ia
  Supernovae Using the Full 5 yr Data Set}},  {\em Astrophys. J. Lett.} {\bf
  973} (2024), no.~1 L14, [\href{http://arxiv.org/abs/2401.02929}{{\tt
  arXiv:2401.02929}}].

\bibitem{DESI:2025fii}
{\bf DESI} Collaboration, K.~Lodha et~al., {\it {Extended Dark Energy analysis
  using DESI DR2 BAO measurements}},
  \href{http://arxiv.org/abs/2503.14743}{{\tt arXiv:2503.14743}}.

\bibitem{Nesseris:2025lke}
S.~Nesseris, Y.~Akrami, and G.~D. Starkman, {\it {To CPL, or not to CPL? What
  we have not learned about the dark energy equation of state}},
  \href{http://arxiv.org/abs/2503.22529}{{\tt arXiv:2503.22529}}.

\bibitem{Cortes:2025joz}
M.~Cort{\^e}s and A.~R. Liddle, {\it {On DESI's DR2 exclusion of
  $\Lambda$CDM}},  \href{http://arxiv.org/abs/2504.15336}{{\tt
  arXiv:2504.15336}}.

\bibitem{Dinda:2025iaq}
B.~R. Dinda and R.~Maartens, {\it {Physical vs phantom dark energy after DESI:
  thawing quintessence in a curved background}},  {\em Mon. Not. Roy. Astron.
  Soc.} {\bf 542} (2025) L31--L35, [\href{http://arxiv.org/abs/2504.15190}{{\tt
  arXiv:2504.15190}}].

\bibitem{Chen:2025mlf}
S.-F. Chen and M.~Zaldarriaga, {\it {It's all Ok: curvature in light of BAO
  from DESI DR2}},  {\em JCAP} {\bf 08} (2025) 014,
  [\href{http://arxiv.org/abs/2505.00659}{{\tt arXiv:2505.00659}}].

\bibitem{Guedezounme:2025wav}
S.~L. Guedezounme, B.~R. Dinda, and R.~Maartens, {\it {Phantom crossing or dark
  interaction?}},  \href{http://arxiv.org/abs/2507.18274}{{\tt
  arXiv:2507.18274}}.

\bibitem{Leauthaud:2025azz}
A.~Leauthaud and A.~Riess, {\it {Looking beyond lambda}},  {\em Nature Astron.}
  {\bf 9} (8, 2025) 1123, [\href{http://arxiv.org/abs/2509.00359}{{\tt
  arXiv:2509.00359}}].

\bibitem{Wolf:2025acj}
W.~J. Wolf, P.~G. Ferreira, and C.~Garc{\'\i}a-Garc{\'\i}a, {\it {Cosmological
  constraints on Galileon dark energy with broken shift symmetry}},
  \href{http://arxiv.org/abs/2509.17586}{{\tt arXiv:2509.17586}}.

\bibitem{Li:2024qso}
T.-N. Li, P.-J. Wu, G.-H. Du, S.-J. Jin, H.-L. Li, J.-F. Zhang, and X.~Zhang,
  {\it {Constraints on Interacting Dark Energy Models from the DESI Baryon
  Acoustic Oscillation and DES Supernovae Data}},  {\em Astrophys. J.} {\bf
  976} (2024), no.~1 1, [\href{http://arxiv.org/abs/2407.14934}{{\tt
  arXiv:2407.14934}}].

\bibitem{Li:2025owk}
T.-N. Li, G.-H. Du, Y.-H. Li, P.-J. Wu, S.-J. Jin, J.-F. Zhang, and X.~Zhang,
  {\it {Probing the sign-changeable interaction between dark energy and dark
  matter with DESI baryon acoustic oscillations and DES supernovae data}},
  {\em Sci. China Phys. Mech. Astron.} {\bf 69} (2026), no.~1 210413,
  [\href{http://arxiv.org/abs/2501.07361}{{\tt arXiv:2501.07361}}].

\bibitem{RoyChoudhury:2024wri}
S.~Roy~Choudhury and T.~Okumura, {\it {Updated Cosmological Constraints in
  Extended Parameter Space with Planck PR4, DESI Baryon Acoustic Oscillations,
  and Supernovae: Dynamical Dark Energy, Neutrino Masses, Lensing Anomaly, and
  the Hubble Tension}},  {\em Astrophys. J. Lett.} {\bf 976} (2024), no.~1 L11,
  [\href{http://arxiv.org/abs/2409.13022}{{\tt arXiv:2409.13022}}].

\bibitem{RoyChoudhury:2025dhe}
S.~Roy~Choudhury, {\it {Cosmology in Extended Parameter Space with DESI Data
  Release 2 Baryon Acoustic Oscillations: A 2{\ensuremath{\sigma}}+ Detection
  of Nonzero Neutrino Masses with an Update on Dynamical Dark Energy and
  Lensing Anomaly}},  {\em Astrophys. J. Lett.} {\bf 986} (2025), no.~2 L31,
  [\href{http://arxiv.org/abs/2504.15340}{{\tt arXiv:2504.15340}}].

\bibitem{Wang:2025znm}
J.-Q. Wang, R.-G. Cai, Z.-K. Guo, and S.-J. Wang, {\it {Resolving the
  Planck-DESI tension by non-minimally coupled quintessence}},
  \href{http://arxiv.org/abs/2508.01759}{{\tt arXiv:2508.01759}}.

\bibitem{Ruchika:2024lgi}
Ruchika, {\it {2D BAO vs 3D BAO: Hints for new physics?}},  {\em Phys. Rev. D}
  {\bf 112} (2025), no.~6 063503, [\href{http://arxiv.org/abs/2406.05453}{{\tt
  arXiv:2406.05453}}].

\bibitem{Hogas:2025ahb}
M.~H{\"o}g{\r{a}}s and E.~M{\"o}rtsell, {\it {Bimetric gravity improves the fit
  to DESI BAO and eases the Hubble tension}},
  \href{http://arxiv.org/abs/2507.03743}{{\tt arXiv:2507.03743}}.

\bibitem{Chaudhary:2025pcc}
H.~Chaudhary, S.~Capozziello, V.~K. Sharma, and G.~Mustafa, {\it {Does DESI DR2
  Challenge {\ensuremath{\Lambda}}CDM Paradigm?}},  {\em Astrophys. J.} {\bf
  992} (2025), no.~2 194, [\href{http://arxiv.org/abs/2507.21607}{{\tt
  arXiv:2507.21607}}].

\bibitem{Chaudhary:2025vzy}
H.~Chaudhary, S.~Capozziello, V.~K. Sharma, I.~G{\'o}mez-Vargas, and
  G.~Mustafa, {\it {Evidence for Evolving Dark Energy from LRG1-2 and Low-$z$
  SNe Ia Data}},  \href{http://arxiv.org/abs/2508.10514}{{\tt
  arXiv:2508.10514}}.

\bibitem{Chaudhary:2025uzr}
H.~Chaudhary, S.~Capozziello, S.~Praharaj, S.~K.~J. Pacif, and G.~Mustafa, {\it
  {Is the LambdaCDM Model in Crisis?}},
  \href{http://arxiv.org/abs/2509.17124}{{\tt arXiv:2509.17124}}.

\bibitem{Mukherjee:2025ytj}
P.~Mukherjee and A.~A. Sen, {\it {New expansion rate anomalies at
  characteristic redshifts geometrically determined using DESI-DR2 BAO and
  DES-SN5YR observations}},  {\em Rept. Prog. Phys.} {\bf 88} (2025), no.~9
  098401, [\href{http://arxiv.org/abs/2505.19083}{{\tt arXiv:2505.19083}}].

\bibitem{Steinhardt:2025znn}
C.~L. Steinhardt, P.~Phillips, and R.~Wojtak, {\it {Dark Energy Constraints and
  Joint Cosmological Inference from Mutually Inconsistent Observations}},
  \href{http://arxiv.org/abs/2504.03829}{{\tt arXiv:2504.03829}}.

\bibitem{Afroz:2025iwo}
S.~Afroz and S.~Mukherjee, {\it {Hint towards inconsistency between BAO and
  Supernovae Dataset: The Evidence of Redshift Evolving Dark Energy from DESI
  DR2 is Absent}},  \href{http://arxiv.org/abs/2504.16868}{{\tt
  arXiv:2504.16868}}.

\bibitem{Favale:2024sdq}
A.~Favale, A.~G{\'o}mez-Valent, and M.~Migliaccio, {\it {Quantification of 2D
  vs 3D BAO tension using SNIa as a redshift interpolator and test of the
  Etherington relation}},  {\em Phys. Lett. B} {\bf 858} (2024) 139027,
  [\href{http://arxiv.org/abs/2405.12142}{{\tt arXiv:2405.12142}}].

\bibitem{Colgain:2024mtg}
E.~O. Colg\'ain and M.~M. Sheikh-Jabbari, {\it {DESI and SNe: Dynamical Dark
  Energy, $\Omega_m$ Tension or Systematics?}},
  \href{http://arxiv.org/abs/2412.12905}{{\tt arXiv:2412.12905}}.

\bibitem{Mukherjee:2025fkf}
P.~Mukherjee and A.~A. Sen, {\it {A New $\sim 5\sigma$ Tension at
  Characteristic Redshift from DESI-DR1 BAO and DES-SN5YR Observations}},
  \href{http://arxiv.org/abs/2503.02880}{{\tt arXiv:2503.02880}}.

\bibitem{Berti:2025phi}
M.~Berti, E.~Bellini, C.~Bonvin, M.~Kunz, M.~Viel, and M.~Zumalacarregui, {\it
  {Reconstructing the dark energy density in light of DESI BAO observations}},
  {\em Phys. Rev. D} {\bf 112} (2025), no.~2 023518,
  [\href{http://arxiv.org/abs/2503.13198}{{\tt arXiv:2503.13198}}].

\bibitem{Scherer:2025esj}
M.~Scherer, M.~A. Sabogal, R.~C. Nunes, and A.~De~Felice, {\it {Challenging the
  {\ensuremath{\Lambda}}CDM model: 5{\ensuremath{\sigma}} evidence for a
  dynamical dark energy late-time transition}},  {\em Phys. Rev. D} {\bf 112}
  (2025), no.~4 043513, [\href{http://arxiv.org/abs/2504.20664}{{\tt
  arXiv:2504.20664}}].

\bibitem{Wang:2025mqz}
Y.~Wang and W.~Lin, {\it {Uncalibrated Cosmic Standards as a Robust Test on
  Late-time Cosmological Models}},  {\em Astrophys. J.} {\bf 989} (2025), no.~1
  120, [\href{http://arxiv.org/abs/2506.04333}{{\tt arXiv:2506.04333}}].

\bibitem{CosmoVerseNetwork:2025alb}
{\bf CosmoVerse Network} Collaboration, E.~Di~Valentino et~al., {\it {The
  CosmoVerse White Paper: Addressing observational tensions in cosmology with
  systematics and fundamental physics}},  {\em Phys. Dark Univ.} {\bf 49}
  (2025) 101965, [\href{http://arxiv.org/abs/2504.01669}{{\tt
  arXiv:2504.01669}}].

\bibitem{Huang:2025som}
L.~Huang, R.-G. Cai, and S.-J. Wang, {\it {The DESI DR1/DR2 evidence for
  dynamical dark energy is biased by low-redshift supernovae}},  {\em Sci.
  China Phys. Mech. Astron.} {\bf 68} (2025), no.~10 100413,
  [\href{http://arxiv.org/abs/2502.04212}{{\tt arXiv:2502.04212}}].

\bibitem{Matthewson:2024ffb}
W.~L. Matthewson and A.~Shafieloo, {\it {Star-crossed labours: checking
  consistency between current supernovae compilations}},  {\em JCAP} {\bf 01}
  (2025) 064, [\href{http://arxiv.org/abs/2409.02550}{{\tt arXiv:2409.02550}}].

\bibitem{Ling:2025lmw}
J.-L. Ling, G.-H. Du, T.-N. Li, J.-F. Zhang, S.-J. Wang, and X.~Zhang, {\it
  {Model-independent cosmological inference after the DESI DR2 data with
  improved inverse distance ladder}},  {\em Phys. Rev. D} {\bf 112} (2025),
  no.~8 083528, [\href{http://arxiv.org/abs/2505.22369}{{\tt
  arXiv:2505.22369}}].

\bibitem{Huang:2024gfw}
L.~Huang, R.-G. Cai, S.-J. Wang, J.-Q. Liu, and Y.-H. Yao, {\it {Narrowing down
  the Hubble tension to the first two rungs of distance ladders}},  {\em Sci.
  China Phys. Mech. Astron.} {\bf 68} (2025), no.~8 280405,
  [\href{http://arxiv.org/abs/2410.06053}{{\tt arXiv:2410.06053}}].

\bibitem{Gialamas:2024lyw}
I.~D. Gialamas, G.~H\"utsi, K.~Kannike, A.~Racioppi, M.~Raidal, M.~Vasar, and
  H.~Veerm\"ae, {\it {Interpreting DESI 2024 BAO: late-time dynamical dark
  energy or a local effect?}},  \href{http://arxiv.org/abs/2406.07533}{{\tt
  arXiv:2406.07533}}.

\bibitem{Lopez-Hernandez:2025lbj}
M.~Lopez-Hernandez, E.~{\'O}. Colg{\'a}in, S.~Pourojaghi, and M.~M.
  Sheikh-Jabbari, {\it {Crosschecking Cosmic Distances from DESI BAO and DES
  SNe Points to Systematics}},  \href{http://arxiv.org/abs/2510.04179}{{\tt
  arXiv:2510.04179}}.

\bibitem{Efstathiou:2024xcq}
G.~Efstathiou, {\it {Evolving dark energy or supernovae systematics?}},  {\em
  Mon. Not. Roy. Astron. Soc.} {\bf 538} (2025), no.~2 875--882,
  [\href{http://arxiv.org/abs/2408.07175}{{\tt arXiv:2408.07175}}].

\bibitem{DES:2025tir}
{\bf DES} Collaboration, M.~Vincenzi et~al., {\it {Comparing the DES-SN5YR and
  Pantheon+ SN cosmology analyses: Investigation based on ''Evolving Dark
  Energy or Supernovae systematics?''}},  {\em Mon. Not. Roy. Astron. Soc.}
  {\bf 541} (2025), no.~3 2585--2593,
  [\href{http://arxiv.org/abs/2501.06664}{{\tt arXiv:2501.06664}}].

\bibitem{Yang:2025qdg}
F.~Yang, X.~Fu, B.~Xu, K.~Zhang, Y.~Huang, and Y.~Yang, {\it {Testing the
  cosmic distance duality relation using Type Ia supernovae and BAO
  observations}},  {\em Eur. Phys. J. C} {\bf 85} (2025), no.~2 186,
  [\href{http://arxiv.org/abs/2502.05417}{{\tt arXiv:2502.05417}}].

\bibitem{Li:2025htp}
T.-N. Li, G.-H. Du, P.-J. Wu, J.-Z. Qi, J.-F. Zhang, and X.~Zhang, {\it
  {Testing the cosmic distance duality relation with baryon acoustic
  oscillations and supernovae data}},
  \href{http://arxiv.org/abs/2507.13811}{{\tt arXiv:2507.13811}}.

\bibitem{Avila:2025sjz}
F.~Avila, F.~Oliveira, C.~Franco, M.~Lopes, R.~Holanda, R.~C. Nunes, and
  A.~Bernui, {\it {Probing the Cosmic Distance Duality Relation via
  Non-Parametric Reconstruction for High Redshifts}},
  \href{http://arxiv.org/abs/2509.07848}{{\tt arXiv:2509.07848}}.

\bibitem{Dinda:2022vmb}
B.~R. Dinda, {\it {Minimal model-dependent constraints on cosmological nuisance
  parameters and cosmic curvature from combinations of cosmological data}},
  {\em Int. J. Mod. Phys. D} {\bf 32} (2023), no.~11 2350079,
  [\href{http://arxiv.org/abs/2209.14639}{{\tt arXiv:2209.14639}}].

\bibitem{Dinda:2024ktd}
B.~R. Dinda and R.~Maartens, {\it {Model-agnostic assessment of dark energy
  after DESI DR1 BAO}},  {\em JCAP} {\bf 01} (2025) 120,
  [\href{http://arxiv.org/abs/2407.17252}{{\tt arXiv:2407.17252}}].

\bibitem{Dinda:2021ffa}
B.~R. Dinda, {\it {Cosmic expansion parametrization: Implication for curvature
  and H0 tension}},  {\em Phys. Rev. D} {\bf 105} (2022), no.~6 063524,
  [\href{http://arxiv.org/abs/2106.02963}{{\tt arXiv:2106.02963}}].

\bibitem{Efstathiou:2021ocp}
G.~Efstathiou, {\it {To H0 or not to H0?}},  {\em Mon. Not. Roy. Astron. Soc.}
  {\bf 505} (2021), no.~3 3866--3872,
  [\href{http://arxiv.org/abs/2103.08723}{{\tt arXiv:2103.08723}}].

\end{thebibliography}\endgroup

\end{document}